\documentclass[prb,twocolumn,preprintnumbers,amsmath,amssymb,showpacs]{revtex4-1}

\usepackage{epsfig,amsfonts}
\usepackage{amsmath}

\def\be{\begin{equation}}
\def\ee{\end{equation}}

\def\bea{\begin{eqnarray}}
\def\eea{\end{eqnarray}}

\def\a{\alpha}
\def\b{\beta}

\def\e{\epsilon}
\def\l{\lambda}

\def\ben{\begin{enumerate}}
\def\een{\end{enumerate}}

\newcommand{\braket}[2]{\langle #1 | #2 \rangle}
\newcommand{\ket}[1]{| #1 \rangle}
\newcommand{\bra}[1]{\langle #1 |}
\def\up{\uparrow}
\def\down{\downarrow}

\begin{document}

\title{Non-equilibrum dynamics in the strongly excited inhomogeneous Dicke model}
\author{Christoph Str\"{a}ter${}^{1,2}$, Oleksandr Tsyplyatyev${}^3$, Alexandre Faribault${}^{4,2}$}
\affiliation{$^1$Max-Planck-Institute for the Physics of Complex Systems, N\"{o}thnitzer Str. 38, 01187 Dresden, Germany}
\affiliation{$^2$Arnold Sommerfeld Centre for Theoretical Physics, Ludwig-Maximilians-Universit\"{a}t, 80333 M\"{u}nchen, Germany}
\affiliation{$^3$Department of Physics \& Astronomy, University of Sheffield, Sheffield S3 7RH, United Kingdom}
\affiliation{$^4$Institute for Theory of Statistical Physics, RWTH Aachen, 52056 Aachen, Germany}

\date{\today}

\begin{abstract}

Using the exact eigenstates of the inhomogeneous Dicke model obtained by numerically solving the Bethe equations, we study the decay of bosonic excitations due to the coupling of the mode to an ensemble of two-level (spin $1/2$) systems. We compare the quantum time-evolution of the bosonic mode population with the mean-field description which, for a few bosons agree up to a relatively long Ehrenfest time. We demonstrate that additional excitations lead to a dramatic shortening of the period of validity of the mean-field analysis. However, even in the limit where the number of bosons equal the number of spins, the initial instability remains adequately described by the mean-field approach leading to a finite, albeit short, Ehrenfest time. Through finite size analysis, we also present indications that the mean-field approach could still provide an adequate description for thermodynamically large systems even at long times. However, for mesoscopic systems one cannot expect it to capture the behavior beyond the initial decay stage in the limit of an extremely large number of excitations.

\end{abstract}


\maketitle



\section{Introduction}

Since they constitute one of the broad classes of proposed physical realizations of quantum computing devices \cite{qucom1,qucom2,qucom3,qucom4}, coherently interacting light-matter systems have received lately a considerable level of attention. The interest has been further enhanced by relatively recent progress in a variety of systems ranging from polaritons in quantum wells \cite{excitons1,excitons2, eastham} to semiconductor quantum dots \cite{dots1,dots2,phillips} which nowadays make it possible to realize solid-state based quantum systems which couple to a single photon eigenmode of optical microcavities. At the same time the many-body effects in these systems are becoming increasingly important in the context of engineering of semiconductor lasers\cite{QD_lasers1,QD_lasers2,QD_lasers3,QD_lasers4} motivated by potentially a great enhancement in performance of designs in which quantum dots serve as an active medium \cite{arakawa}.

While an ideal system of identical two-level (spin 1/2) emitters coupled uniformly to a single light-mode is describable in terms of the Dicke Hamiltonian \cite{dicke,nonRWA}, a more realistic setup would need to include possible inhomogeneities in both the energy splitting of the individual spins and in their respective coupling strengths to the bosonic light mode. In previous studies \cite{oleks_dynamics, oleks_ehrenfest}, the comparison of the quantum and mean-field dynamics of the resulting generalized Dicke model

\bea
	H=\omega b^{\dag}b + \sum_{j=1}^N \epsilon_j S_j^z + \sum_{j=1}^N V_j \left( b^{\dag} S_j^- + S_j^+b\right),
\label{dickehamvj}
\eea

\noindent  was performed by solving the Schr\"{o}dinger equation in the limit of small excitation numbers. Since the total number of excitations $M=b^\dag b + \sum_i \left(S^z_i+\frac{1}{2}\right) $ is conserved, the dimension of the Hilbert space involved in the unitary evolution of the system is drastically reduced. Thus it became possible to solve the explicit time dependence of every quantum amplitude involved.

In this work, we revisit this problem by exploiting the quantum integrability of a certain generalized Dicke model.  Using the algebraic Bethe Ansatz (ABA), one can numerically compute exact eigenstates of the system and study its dynamics rendered trivial by the use of the proper eigenbasis. Additionally, this approach allows, in the strong coupling regime, a drastic truncation of the Hilbert space granting access to relatively large system sizes. On the other hand, integrability imposes the constraint that every spin-like subsystem be uniformly coupled to the bosonic mode and consequently, we study the specific Hamiltonian
\bea
	H=\omega b^{\dag}b + \sum_{j=1}^N \epsilon_j S_j^z + V \sum_{j=1}^N  \left( b^{\dag} S_j^- + S_j^+b\right).
\label{dickeham}
\eea
It is known\cite{oleks_dynamics} that the eigenstates of Eq. (1) are the eigenstates of Eq. (2) with $V=\sqrt{\sum_{j=1}^N V_j^2/N}$, at least\cite{Vj_many} when the number of excitations is small $M\ll N$.

As in Ref. [\onlinecite{oleks_ehrenfest}], this work focuses on the decay of a number of bosonic excitations due to the coupling of the mode to an initially unexcited set of two-level emitters. Such a state could, in principle, be obtained by first preparing the spin system in its ground-state then exciting the bosonic mode via an external radiation pulse. For small number of excitations, a crossover between two distinct regimes was found. At weak coupling, the bosonic occupation number $\left<b^\dag b\right>$ undergoes an exponential decay and at strong enough coupling, the occupation undergoes non-decaying periodic oscillation with frequency which is enhanced by the Dicke supperradiance effect. This is a dynamical counterpart of the Hepp and Lieb superradiance quantum phase transition.\cite{HeppLieb}

Starting from a large number of excitations, for weak coupling strength, the dimension of the necessary Hilbert space severely limits the system sizes treatable using the ABA. Still, for very small systems, we do find rapidly decaying short-time dynamics as in the mean-field treatment. However, our capacity to make quantitative comparisons with the mean-field approach is hindered since its validity is necessarily limited to large systems. Consequently, the bulk of our results focuses on the strong coupling regime, where heavy truncation of the Hilbert space allows a nearly exact treatment of larger systems. In this case, for a small number of initial excitations, the spectrum obtained in the full quantum treatment is characterized by set of equally spaced frequencies, leading to nearly periodic real-time dynamics. The mean-field approach, which also leads to periodical oscillating bosonic populations, then reproduces the quantum dynamics up to some relatively long Ehrenfest time at which both solutions start to differ significantly. 

When the number of initial excitations becomes of the order of the system size, the spectrum shows strong deviations from a harmonic progression no longer reproducing the periodic oscillations obtained in mean-field. Nonetheless, we find that the mean-field approach remains valid for the very short-time dynamics of the system leading to a finite, but considerably shortened, Ehrenfest time.

The crossover to the regime of periodic oscillations occurs when the superradiantly enhanced coupling with the boson mode becomes larger than the bandwidth of the spin energy splittings. For example, this effect can  manifest itself  as a suppression of the inhomogeneous broadening in a system of self-assembled quantum dots in a zero dimensional cavity. A realistic high in-plane density of InGaAs dots\cite{InGaAsdots} $\sim 5\cdot 10^{10}\;\textrm{cm}^{-2}$ will enhance coupling of a single dot to a photonic crystal cavity,\cite{dots2} a micropillar,\cite{Lemaitre} or a concave microcavity,\cite{Warburton} which is ($\sim 3-100$ $\mu$eV), by two to three orders of magnitude in the optical domain\cite{Nestimate} that is well inside or even above the range of natural bandwidths of ensembles of such dots\cite{bandwidth} ($\sim 5-50$ meV). Thus for sufficiently high densities  a spectrally broadened ensemble of the self-assembled dots will exchange all of its excitations with a boson mode  without a decay, at least on a finite time scale, in the same way as an ideal atomic system without any broadening, making it suitable to engineer a high power semiconductor laser. 

When such a system is in the strong coupling regime, the effect of quantum fluctuations could be observed in a direct time-resolved measurement at different excitation powers. At low to intermediate number of excitations the many Rabi oscillations would be visible without any significant decay. However, when the number of excitations reaches $\sim 80$\% of the number of spins, a strong decay on a time scale of a single period would appear due to quantum fluctuations. The dependence of this effect on the number of spins can be used to discriminate it from other sources of relaxation that occur at large probing powers such as charge or phonon fluctuations.

The paper is organized as follows. Section \ref{bethe} describes the exact solution (ABA) and the numerical techniques used to exploit it in order to compute the non-equilibrium dynamics we are interested in. Section \ref{class} covers the mean-field approach to the same problem. The resulting behavior which stems from both descriptions are then compared and analyzed in Section \ref{results} where both the spectrums and the real-time dynamics are studied.


\section{The Exact Solution}
\label{bethe}

In the following analysis, we use  $\epsilon_i$ (single spin excitation energies) which are uniformly distributed $\e_{i+1}-\e_i=\epsilon_d N/(N-1)$ within a band of total width $\Delta=\e_N-\e_1=N$, here $\epsilon_d$ is a ``level spacing'' for spins. This width $\Delta$ will serve as a natural energy scale of the problem. Furthermore, we introduce the Rabi frequency $\Omega=V\sqrt{N}$, which is the oscillation frequency for bosons in the case of equal splittings $\e_i\equiv\e$. The ratio $\Omega/\Delta$ is thus a dimensionless parameter that we will use to specify the coupling strength. 


\subsection{Constructing Eigenstates}

We exactly solve the Dicke Hamiltonian (\ref{dickeham}) using the method introduced in Refs. [\onlinecite{methods}] and [\onlinecite{methods2}]. Hereby, we exploit the quantum integrability of the Dicke Hamiltonian, which was proven in [\onlinecite{dicke_proof}]. Unnormalized eigenstates can thus be constructed by creating $M$ pseudo-particles,
\bea
	\prod_{\a=1}^M \mathrm{S}^+(\l_{\a}) \ket{0;\down\ldots\down},
\label{eigenstates}
\eea	
 on the vacuum state $\ket{0;\down\ldots\down}$ which contains no bosons and has all spins in their lowest energy states. For the Dicke model (\ref{dickeham}) the creation operator takes the form
\bea
	\mathrm{S}^+(\l_{\a})=b^{\dag}+\sum_{j=1}^N \frac{V}{\l_{\a}-\e_j}S_j^+.
\label{creationop}
\eea
States of the form (\ref{eigenstates}) become eigenstates of the Dicke Hamiltonian when the $M$ rapidities $\l_{a}$ fulfill the $M$ Bethe equations
\bea
	\sum_{\substack{\b=1 \\ \b\neq\a}}^M \frac{V}{\l_{\a}-\l_{\b}}=\frac{\omega}{2V}-\frac{\l_{\a}}{2V}+ \frac{1}{2} \sum_{j=1}^N\frac{V}{\l_{\a}-\e_j},
\label{betheequations}
\eea

\noindent which one can obtain from a straightforward application of the Hamiltonian (\ref{dickeham}) to a general state (\ref{eigenstates}) with unspecified rapidities $\lambda_\a$  \cite{poorman}.

Being completely defined by a set of $M$ rapidities $\{\l\}$, we denote the (unnormalized) eigenstates by $\ket{\{\l\}}$. The corresponding eigenenergies are simply given by
\bea
	E_{\{\l\}}=\sum_{\a=1}^M \l_{\a} .
\eea

Due to numerical instabilities when trying to solve the Bethe ansatz Eqs. (\ref{betheequations}) directly (see for example [\onlinecite{faribaultjmp}]), we use the change of variables proposed in [\onlinecite{methods},\onlinecite{methods2}]. First we introduce the complex polynomial
\bea
	\Gamma(z)=\sum_{\a=1}^M\frac{1}{z-\l_{\a}},
\eea
\noindent which we evaluate at the Zeeman splittings $z = \e_j$ to obtain $N$ new variables $K_j=V^2\Gamma(\e_j)$. For a set of rapidities that are a solution to (\ref{betheequations}) one can show that the corresponding new variables obey a set of quadratic equations
\bea	
	V^2\sum_{\substack{i=1 \\ i\neq j}}^N \frac{K_i-K_j}{\e_i-\e_j} + V^2M = K_j \left(\e_j-\omega\right) + K_j^2.
\label{bethesubstituted}
\eea		
Note that instead of $M$, we now have $N$ equations. As long as $M\le N$, these equations are equivalent to the Bethe ansatz Eqs. (\ref{betheequations}), but they lack the numerical problems mentioned before. For every quasi-particle number $M$, which is conserved in time, the equations allow for several solutions $\{K\}$, all of them in one to one correspondence to a given set of rapidities $\{\l\}$ and thus to a single eigenstate of the system.

At $V=0$, the eigenstates of the system are obviously Fock states with a definite number of spin excitations and bosons. For example, for $N=2$ and $M=2$ the Fock states are $\ket{2;\down\down}$, $\ket{1;\up\down}$, $\ket{1;\down\up}$ and $\ket{0;\up\up}$ where, in this notation, the number counts the bosonic excitations whereas the arrows represent the spin states. From the Bethe Eqs. (\ref{betheequations}) it immediately follows that the set of rapidities $\{\l\}$ of a Fock state consists of one $\l_{\a}=\omega$ for every boson and one $\l_{\a}=\e_i$ if the $i$th spin is excited. This leads to $K_i = \omega-\e_i$ if the $i$th spin is excited and $K_i = 0$ if it is not. The total number of excitations $M$ then differentiates between states with identical spin configurations but different number of bosons. For example, the state $\ket{1;\up\down}$ has $\{\l\}=\{\omega,\e_1\}$ and $\{K\}=\{0,\omega-\e_1\}$.  


For a desired final coupling $V=V_f$, we obtain the solutions $\{K\}$ by deforming the solutions at $V=0$ by a stepwise increasing of $V$. As detailed in Ref. [\onlinecite{methods}], one can compute easily the $n$th first derivatives $\frac{\partial^n K_i(V)}{\partial V^n}$. This provides an good initial guess for the solution at $V+dV$ using the values $K_i(V)$ through the truncated Taylor expansion. One can the refine this guess using a simple iterative Newton-Raphson algorithm applied to the quadratic system of Eqs. (\ref{bethesubstituted}). The process can then be repeated until the target coupling value $V_f$ is reached.

Since the rapidities themselves are used to calculate physical quantities, we need to extract the set of $M$ values $\{\l\}$ corresponding to a given set $\{K\}$. In this work it is achieved by using the fact that 

\bea
 \Gamma (z)=\frac{Q(z)}{Q'(z)},
\eea

\noindent with 
\bea
	Q(z)=\prod_{\beta=1}^M(z-\lambda_{\beta})=\sum_{\alpha=0}^M Q_{\alpha} z^{M-\alpha}.
\eea

The coefficients of this polynomial 
\bea
	Q_{\alpha}=(-1)^{\a}\sum_{\substack{k_i=1\\k_i\neq k_j}}^M \lambda_{k_1}\cdots\lambda_{k_{\alpha}}
\eea	
are  the elementary symmetric polynomials which can be found by solving the linear system: \begin{equation}
	\sum_{\alpha=1}^M \left[(M-\alpha)  \epsilon_j^{M-\alpha-1}-\frac{ K_j}{V^2}\epsilon_j^{M-\alpha}\right] Q_{\a}=  \frac{ K_j}{V^2}\epsilon_j^M-M\epsilon_j^{M-1}.
\label{setofeq}
\end{equation}

The last task is to find the rapidities $\{\l\}$ as the roots of the polynomial $Q(z)$ by its coefficients $Q_{\a}$, which one can do using a number of root finding algorithms. Using this method, we can, in principle, calculate the rapidities $\{\l\}$ characterizing every single eigenstate of the system.


\subsection{Obtaining the Bosonic Occupation Number}
\label{subsec:bosonocc}

Eventually, we are interested in the time dependent bosonic occupation number $\langle b^{\dag}b\rangle(t)=\bra{\psi(t)}b^{\dag}b\ket{\psi(t)}$ with initial state $\ket{M;\downarrow\ldots\downarrow}$. Expanding in the normalized eigenbasis $\ket{\phi_i}=\ket{\{\l\}_i}/\sqrt{\braket{\{\l\}_i}{\{\l\}_i}}$, we obtain
\bea
	\bra{\psi(t)}b^{\dag}b\ket{\psi(t)}=&&
	\sum_{i,j=1}^d\braket{M;\downarrow\ldots\downarrow}{\phi_i} \bra{\phi_i}b^{\dag}b\ket{\phi_j}\nonumber\\
	&&\braket{\phi_j}{M;\downarrow\ldots\downarrow}\ e^{\frac{i}{\hbar} (E_i-E_j) t},
\label{bosonocc}
\eea
where we denote by $d$ the dimension of the Hilbert space. The matrix elements occurring in (\ref{bosonocc}), as well as the norm of the eigenstates $\braket{\{\l\}_i}{\{\l\}_i}$ can be computed using Slavnov's formula [\onlinecite{slavnov}]. Provided the set $\{\mu\}$ fulfills the Bethe ansatz Eqs. (\ref{betheequations}), one can write its overlap with a generic state of the form (\ref{eigenstates}) as the determinant of an $M$ by $M$ matrix
\bea
	\braket{\{\mu\}}{\{\lambda\}}= \frac{\prod_{k\neq l}^M (\lambda_l-\mu_k)}{\prod_{k>l}(\lambda_k-\lambda_l)\prod_{k<l} (\mu_k-\mu_l)}\det{G},\quad 
\eea
with
\begin{multline}
	G_{\a,\beta}= \left( \omega - \l_{\beta} + \sum_{j=1}^{N} \frac{V^2}{\l_{\beta}-\e_j} + \right. \\ \left. \sum_{\gamma\neq\alpha} \frac{2V^2}{\mu_{\gamma}-\l_{\beta}}\right) \frac{\l_{\beta}-\mu_{\beta}}{(\l_{\beta}-\mu_{\alpha})^2}.
\end{multline}
For the norms of eigenstates we hence obtain, in the limit $\{\mu\}\rightarrow\{\l\}$,
\bea
	\braket{\{\l\}}{\{\l\}}	&=&\det(W),
\eea
where
\begin{equation}
	W_{\alpha\beta}= \frac{2V^2}{(\lambda_{\beta}-\lambda_{\alpha})^2}.
\label{norms}
\end{equation}

In a way similar to [\onlinecite{links}], we can furthermore derive a single determinant expression for the form factor appearing in Eq. (\ref{bosonocc}):
\bea
	\bra{\{\mu\}}&&b^{\dag}b\ket{\{\lambda\}}=\nonumber\\
	  && \frac{\prod_{k\neq l}^M (\lambda_l-\mu_k)}{\prod_{k>l}(\lambda_k-\lambda_l)\prod_{k<l} (\mu_k-\mu_l)} \det(G+Q),
\eea
with the $M\times M$ matrices $Q$ defined by
\bea
	Q_{\alpha\beta}=\prod_{l\neq \beta}\frac{ (\l_{\beta}-\lambda_l)}{ (\l_{\beta}-\mu_l)}.
\eea

On the other hand, since $\ket{M;\downarrow\ldots\downarrow}=1/\sqrt{M!}\ b^{\dag M}\ket{0;\downarrow\ldots\downarrow}$, we can rewrite the overlaps between eigenstates and the initial state by noting that only the bosonic parts of the eigenstates (\ref{creationop}) do not vanish in this projection 
\bea
	\braket{M;\downarrow\ldots\downarrow}{\phi_i}&=&\sqrt{\frac{M!}{\braket{\{\lambda\}}{\{\lambda\}}}}.
\label{projection}
\eea
Hence, we are left with the norms of the eigenstates $\braket{\{\l\}}{\{\l\}}$ which are computed by (\ref{norms}). In this fashion, we are able to compute the bosonic occupation $(\ref{bosonocc})$ fully in terms of the rapidities of all eigenstates.


\subsection{Hilbert Space Truncation}

Although every term in Eq. (\ref{bosonocc}) can be easily computed, the double sum over the complete Hilbert space remains remarkably large. For a system with $O(10)$ spins this sum already becomes impossible to perform fully. In this work we rely on the fact that, at weak enough ($\Omega\ll \Delta/N $) or strong enough ($\Omega\gg \Delta$) coupling, the main contributions to (\ref{bosonocc}) comes from only a small number of eigenstates. 

The truncation scheme for very small coupling works as follows. For $V=0$, the initial state $\ket{M;\down\ldots\down}$ is an eigenstate itself and therefore the only relevant state for the given scenario. Perturbation theory then provides a natural hierarchy such that states where a single excitation is swapped from the bosonic mode to a spin, are the most relevant ones. Therefore, keeping only states with a single spin excitation, for example the state $\ket{M-1;\up\down\ldots\down}$, would lead to a large contribution. One could then add two spin-excitations states and so on. 


At strong coupling, considering Eqs. (\ref{eigenstates}) and (\ref{creationop}) we see that, as $V \to \infty$, any finite rapidity leads to an excitation (created by $\mathrm{S}^+(\lambda)$) which exclusively affects the spin sector. On the other hand, any rapidity, which diverges when $V\rightarrow \infty$, creates an excitation that significantly populates the bosonic mode. When looking at the projection of any eigenstate onto the purely bosonic initial state Eq. (\ref{projection}), we can infer that, at large enough couplings, only the eigenstates for which every one of the $M$ rapidities diverge will lead to significant overlaps. 
Since the form factors $\bra{\phi_i}b^{\dag}b\ket{\phi_j}$ in (\ref{bosonocc}) are bounded by $M$, 
only the eigenstates with all rapidities diverging are needed in the $V\to \infty$ limit. At finite but large $V$, similar argument can be used to show that states with $M-1$ diverging rapidities are the first ones to become important, and so on. This provides a rough ordering of the relative contribution to the sum of different classes of eigenstates. 

Having identified the heavily contributing states, 
 we can truncate the sum in Eq. (\ref{bosonocc}) to a smaller dimension $\tilde{d}$ using only a subset of the most important eigenstates. 
 We estimate the validity of the  truncated sum by computing the sum rule
\bea
	\Sigma(\tilde{d})=\sum_{i=1}^{\tilde{d}} \left|\braket{M;\down\ldots\down}{\phi_i}\right|^2.
	\label{sumrule}
\eea
For the complete sum $\tilde{d}=d$ we have $\Sigma(\tilde{d})=1$ since we are projecting the initial states on the complete eigenbasis. In view of bounding the total error $\delta$ in the time evolution for arbitrary times, we require at least $\Sigma_T(\tilde{d})\ge 1- \delta$. While the factor $\bra{\phi_i}b^{\dag}b\ket{\phi_j}$ in Eq. (\ref{bosonocc}) is different for different eigenstates, it is of the same order of magnitude for  eigenstates containing the same number of divergent rapidities, which can be seen from Eq. (\ref{creationop}). Therefore, the saturation of the sum rule is a clear indication of the error generated by the truncation. Nevertheless, we crosscheck the validity of the truncated sum by checking if $\langle b^{\dag}b\rangle(0)/M> 1-\delta $ at $t=0$. Due to the trivial time evolution in the true eigenbasis, the absolute error should remain bounded by this initial value.



\section{Mean-field analysis}
\label{class}

In this section we study the dynamics of the model Eq. (\ref{dickeham}) using the mean-field approximation. We derive the Hamilton equations of motion in terms of the expectation values of the boson and spin operators, and solve them for the initial conditions from Sec. \ref{subsec:bosonocc} for arbitrary $M$.

Heisenberg equations of motion for quantum operators are obtained from  Eq. (\ref{dickeham}) by use of commutation relations, e.g. $\dot{b}=i \left[ H,b \right]$. The complete set of equations of evolution for the boson and spin operators is
\begin{eqnarray}
	\dot{\mathbf{S}}_{j} & = & \mathbf{B}_{j}\times\mathbf{S}_{j}\label{eq:QEOM1}\\
	\dot{b} & = & -iV\sum_{j=1}^{N}S_{j}^{-},
\label{eq:QEOM2}
\end{eqnarray}
where $z$ and in-plane components of vector $\mathbf{B}_j=\left( 2 V b_x, 2 V b_y, \epsilon_j\right)$ are a single spin splitting and boson operators; $b_x+i b_y=b^\dagger$ and $b_x-i b_y=b$. When the harmonic oscillator is highly excited the boson operator can be approximated by a time-dependent c-number $\langle b\rangle= a$ which makes the systems of Eqs. (\ref{eq:QEOM1}, \ref{eq:QEOM2}) linear in operators. Here $\langle \dots \rangle$ is the time-dependent quantum mechanical expectation value. Averaging the linearized equations over an initial state we obtained the dynamical mean-field equations,
\begin{eqnarray}
\dot{\mathbf{C}}_{j} & = & \mathbf{B}_{j}\times\mathbf{C}_{j}\label{eq:CEOM1}\\
\dot{a} & = & V\sum_{j=1}^{N}C_{j}^{-},\label{eq:CEOM2}
\end{eqnarray}
where $\mathbf{C}_j=\langle S_j \rangle$ is a set of $N$ vectors of length $\left| C_j \right|=1/2$ and $\mathbf{B}_j=\left( 2 V a_x, 2 V a_y, \epsilon_j\right)$; here $C^-_j=C^x_j-iC^y_j$ and $a=a_x-i a_y$. 

Alternatively,  Eqs. (\ref{eq:CEOM1}) and (\ref{eq:CEOM2}) can  be derived using Dirac's analogy for dynamical variables: Commutation relations between quantum operators correspond to Poisson brackets between classical degrees of freedom, $[,]\rightarrow -i[,]_{cl}$. In the model Eq. (\ref{dickeham}) spin operators $\mathbf{S}_j$ can be associated with classical vectors $\mathbf{C}_j$ and the boson operator $b$ with a classical field $a$. By analogy, the Poisson brackets between the classical variables correspond to an angular momentum $\left[C^{\alpha},C^{\beta}\right]_{cl}=-\epsilon_{\alpha\beta\gamma}C^{\gamma}$ and a boson field $\left[a,a^{*}\right]_{cl}=i$ commutation relations.  The Hamilton equations of motion for such a classical model are equivalent to the mean-field approximation. In this way Eqs. (\ref{eq:CEOM1}, \ref{eq:CEOM2}) can be interpreted as the  classical limit of Eqs. (\ref{eq:QEOM1}, \ref{eq:QEOM2}). 

In Sec. \ref{bethe} we wrote down explicit expressions for the quantum dynamics of the initial state $\ket{M;\down\ldots\down}$. The initial condition for the dynamical mean-field equations, which corresponds to this state, is all spins down, $\mathbf{C}_{j}\left(0\right)=\left(0,0,-1/2\right)$,
and a finite amplitude of the bosonic field, $a\left(0\right)=\sqrt{M}$. Below, we solve  set of differential Eqs. (\ref{eq:CEOM1}, \ref{eq:CEOM2}) for the evolution of  classical variables with this initial condition.

For a small number of excitations in a system with many spins, $M\ll N$, these equations were solved in [\onlinecite{oleks_ehrenfest}]. Using the approximation $C_{j}^{z}\left(t\right)\approx-1/2$, the equation
for $C_{j}^{z}\left(t\right)$ drops out from Eq. (\ref{eq:CEOM1})
and the remaining system of equation is harmonic. In the regime $\Omega\gg\Delta$ this gives the following solution for the bosonic mode 
\begin{equation}
	a\left(t\right)=\sqrt{M}\cos\left(V\sqrt{N}t\right).
\label{eq:harmonic}
\end{equation}
The period of the oscillatory function in this limit is $T=2\pi/\left(V\sqrt{N}\right)$,
see Figure \ref{fig:time_evolution} for $M=10$. In the following will use the Rabi frequency $\Omega=V\sqrt{N}$ which was introduced in Sec. II.

\begin{figure}[t]
     \center{
      \includegraphics[width=0.95\columnwidth]{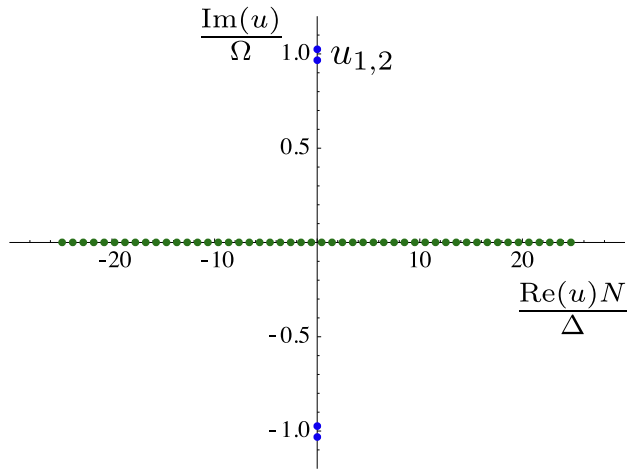}
      }
  \caption{Numerical solution of the equation $\sqrt{L^2(u)}=0$ with $\mathbf{L}(u)$ from Eq. (\ref{eq:lax_vector}) for $N=50$, $M=N$, and $\Omega/\Delta=10$. The roots that merge into a continuous line are plotted in green and two pairs of discreet roots are plotted in blue.}
  \label{fig:L_roots}
\end{figure}

For arbitrary $M$ the solution to Eq. (\ref{eq:CEOM1},
\ref{eq:CEOM2}) are hyperelliptic functions \cite{fermi_bose}. Here we will not consider the full analytic form
of general solution but will use only the spectral analysis developed
in Ref. [\onlinecite{relaxation}]. Introducing the following vector function
(Lax vector) of an auxiliary parameter $u$,
\begin{equation}
\mathbf{L}\left(u\right)=\left(\begin{array}{c}
\frac{a_{x}\left(0\right)}{V}\\
\frac{a_{y}\left(0\right)}{V}\\
\frac{u}{2V^{2}}
\end{array}\right)+\sum_{j}\frac{\mathbf{C}_{j}\left(0\right)}{u-\left(\epsilon_{j}-\omega\right)},\label{eq:lax_vector}
\end{equation}
the frequency spectrum is related to the roots of the equation $\sqrt{L^{2}\left(u\right)}=0$.
We analyze them numerically in the limit $N\gg1$, and find that,
in the high coupling regime $\Omega\gg\Delta$, all
roots merge into a continuous line except two complex conjugated pairs, see example in Figure \ref{fig:L_roots}.
Note that all coefficients of the polynomial $L^{2}\left(u\right)$
are real thus every complex root has a complex conjugated partner.
The dynamical variables that correspond to the continuous band form
a decay part of the solution and the two discreet frequencies give
an oscillating part that we will be interested in. The discreet roots
can be found by turning the summation over $j$ in Eq. (\ref{eq:lax_vector})
into an integral, $\sum_{j}\rightarrow\frac{N}{\Delta}\int_{\omega-\Delta/2}^{\omega+\Delta/2}d\epsilon$.
Then, the equation $\sqrt{L^{2}\left(u\right)}=0$ turns into 
\begin{equation}
\pm2iga\left(0\right)=u-\frac{V^{2}N}{\Delta}\ln\left(\frac{u+\Delta/2}{u-\Delta/2}\right),\label{eq:lax2}
\end{equation}
where  the total width of splittings $\Delta=\epsilon_{N}-\epsilon_1$ is finite. 

In the limit   $M=N$, opposite to $M\ll N$, the roots of Eq. (\ref{eq:lax2})
have zero real part. Parametrizing the roots as $u=iu_{0}\Delta/2$
we obtain

\begin{equation}
\frac{\pm4Va\left(0\right)}{\Delta}=u_{0}+\frac{2V^{2}N}{\Delta^{2}}\left(\pi-2\tan^{-1}u_{0}\right).
\end{equation}
Then a $1/u_{0}$-expansion gives the imaginary parts as $u_{1,2}=u_{0}\Delta/2=\pm\left(\frac{V\sqrt{N}}{2}\pm\frac{\Delta}{2\sqrt{3}}\right)$.

In the case of two discreet roots the hyperelliptic function of many
variables reduces to an elliptic function of only one variable \cite{relaxation} which corresponds to an effective model of a single collective spin coupled to a boson. Following the procedure of constructing a 1-spin
solution in Ref. [\onlinecite{fermi_bose}] we write 
\begin{equation}
\dot{u}=-i\sqrt{Q_{4}\left(u\right)}\label{eq:diff_u_t}
\end{equation}
\begin{equation}
\dot{a}=iau\label{eq:diff_a_t}
\end{equation}
where $Q_{4}\left(u\right)=\left(u^{2}+u_{1}^{2}\right)\left(u^{2}+u_{2}^{2}\right)$
is a polynomial given by the imaginary roots of Eq. (\ref{eq:lax2}).
Here we choose $u_{2}>u_{1}$. 

The differential Eq. (\ref{eq:diff_u_t}) defines a Jacobi elliptic
function
\begin{equation}
u\left(t\right)=iu_{1}\textrm{\textrm{sn}}\left(\left|u_{2}\right|t-A,k\right),
\end{equation}
where $k=\left|u_{1}/u_{2}\right|$ is the elliptic modulus, and $A$
is an unknown constant of integration. Integrating the second equation
separately for $a$ and $u$ we get 
\begin{equation}
a\left(t\right)=B\left(\frac{\textrm{dn}\left(A\right)-\sqrt{k}\textrm{cn}\left(A\right)}{\textrm{dn}\left(u_{2}t-A\right)-\sqrt{k}\textrm{cn}\left(u_{2}t-A\right)}\right)^{\sqrt{k}},\label{eq:a_t}
\end{equation}
where $B$ is a second constant of integration. 

From the initial conditions, the phase of the oscillation at $t=0$
is $A=0$. The second constant of integration is obtained from the
condition $a\left(t=0\right)=\sqrt{N}$ as $B=\sqrt{N}$. Finally,
expanding the parameters in Eq. (\ref{eq:a_t}) for $\Delta^{2}/\left(V^{2}N\right)\ll1$
we obtain 
\begin{equation}
a\left(t\right)=\frac{\frac{2\Delta}{\sqrt{3}V}}{\textrm{dn}\left(\frac{V\sqrt{N}t}{2}\right)-\sqrt{k}\textrm{cn}\left(\frac{V\sqrt{N}t}{2}\right)}.
\end{equation}
The period of the oscillatory function for this initial condition
is given by the complete elliptic integral of the first kind,
$T=8K\left(k\right)/\left(V\sqrt{N}\right)$, see Fig. \ref{fig:time_evolution} for $M=50$.


\section{Results}
\label{results}

\begin{figure}[h]
      \includegraphics[width=0.255\textwidth, angle=270]{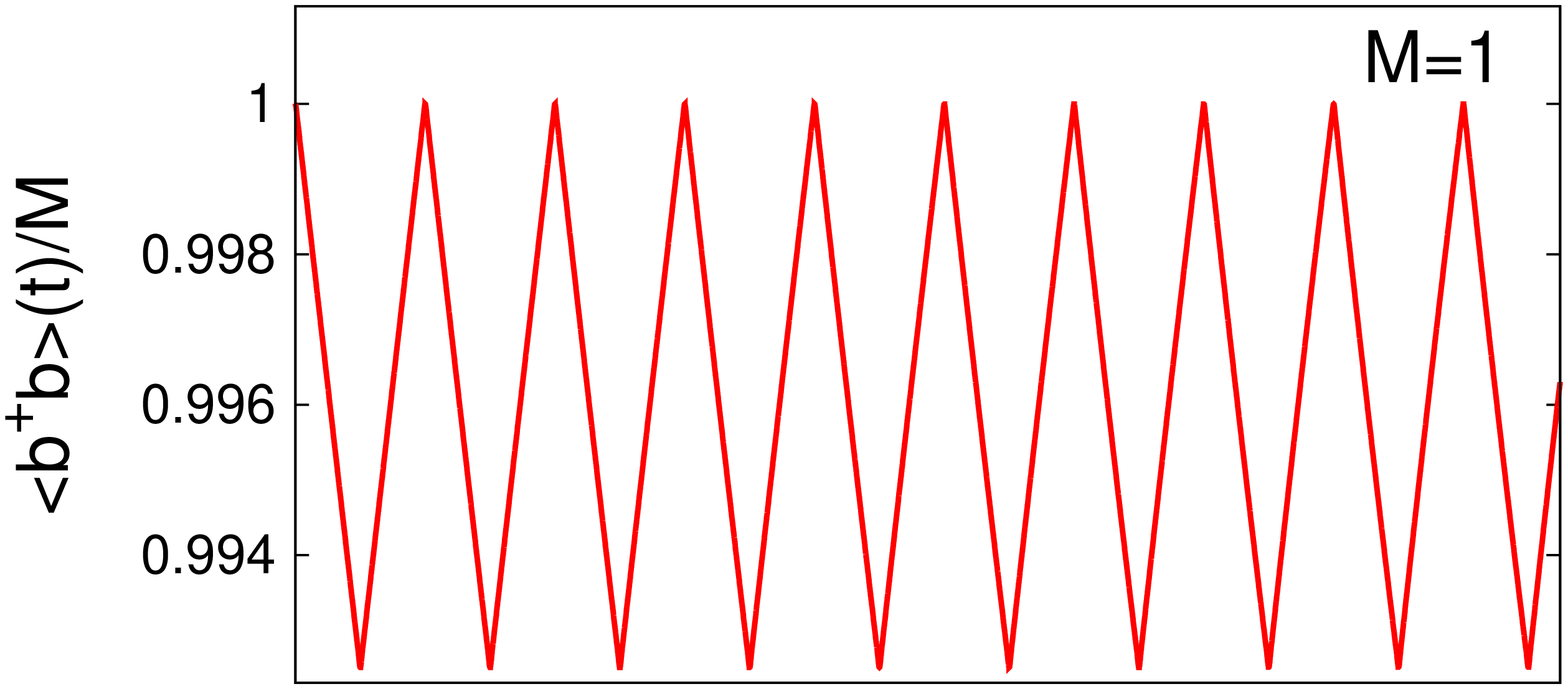}\vspace{-1.05cm}
      \includegraphics[width=0.255\textwidth, angle=270]{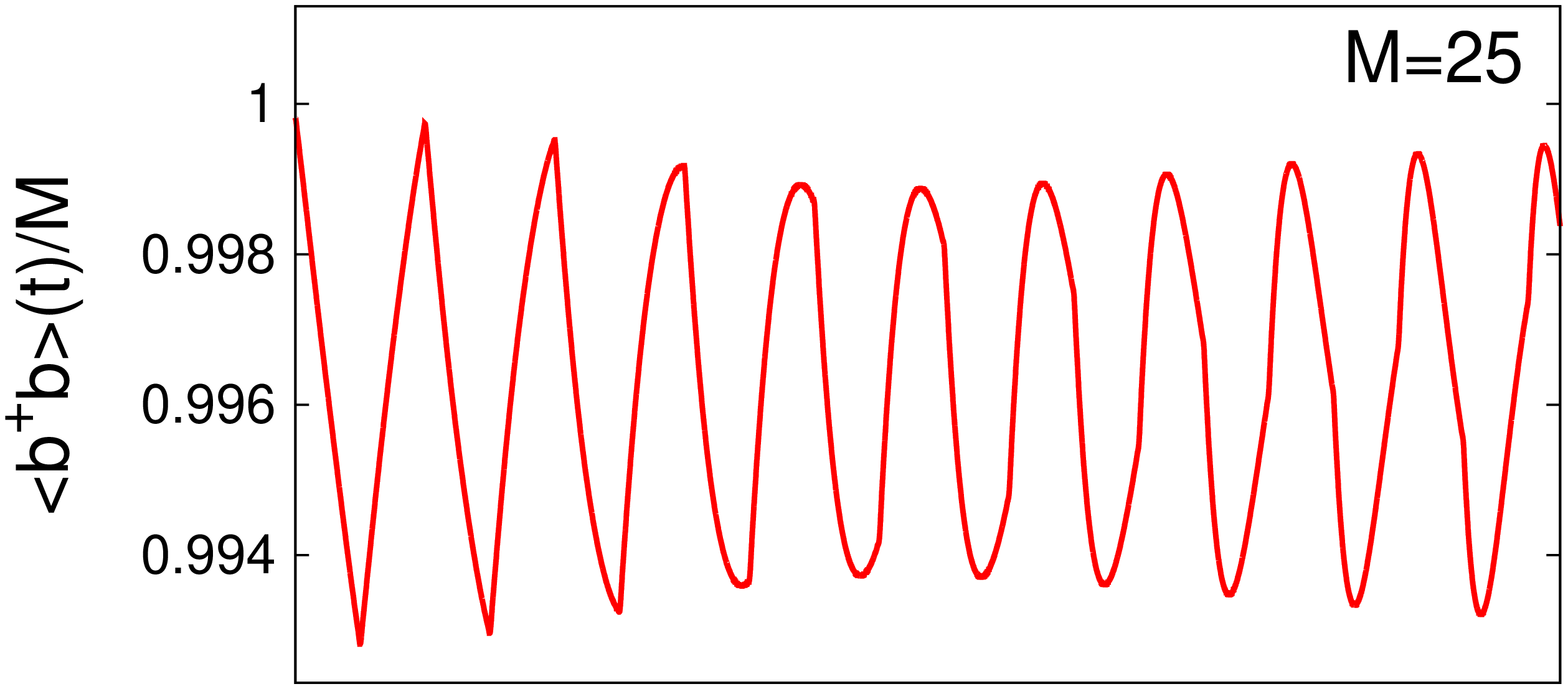}\vspace{-1.05cm}
      \includegraphics[width=0.255\textwidth, angle=270]{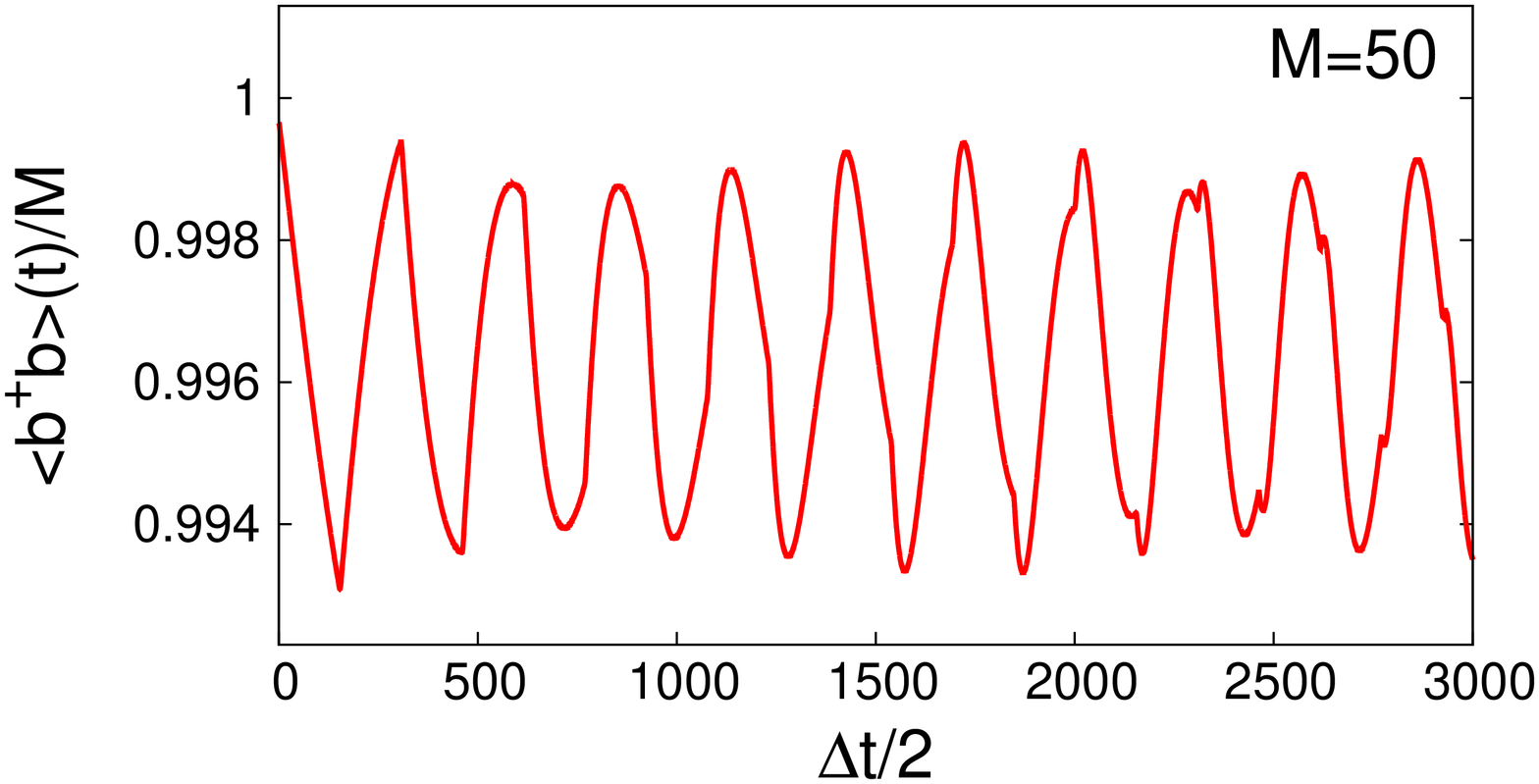}
  \caption{ Time evolution of bosonic occupation number, for $N=50$ spins and $M=1,25,50$ excitations, Coupling strength is $\Omega/\Delta=0.002$. The truncation error is $\delta=0.05 \%$, using only states with a single spin excitation.}
  \label{fig:pert}
\end{figure}

Following the different regimes of interaction strength, we discuss now the effect of a large number of excitations on the time-evolved bosonic occupation. When the interaction strength is very weak, the discreetness of the spin subsystem plays a major role. Indeed, when the Rabi frequency $\Omega$ is smaller than the spacing between $\epsilon_j$s, only spins which are very close to resonance with the bosonic mode are significantly hybridized with the cavity while the rest only plays a weak perturbative role. The resulting dynamics are therefore expected to exhibit non-universal behavior linked to the specific choice of the band's $\epsilon_j$. This results only in weak oscillations around $\left<b^\dag b\right> = M$. A large $M$ (25 and 50 are shown)  does not bring any major qualitative changes to the time evolution, see Fig. \ref{fig:pert}. 

\begin{figure}[h]
      \includegraphics[width=0.27\textwidth, angle=270]{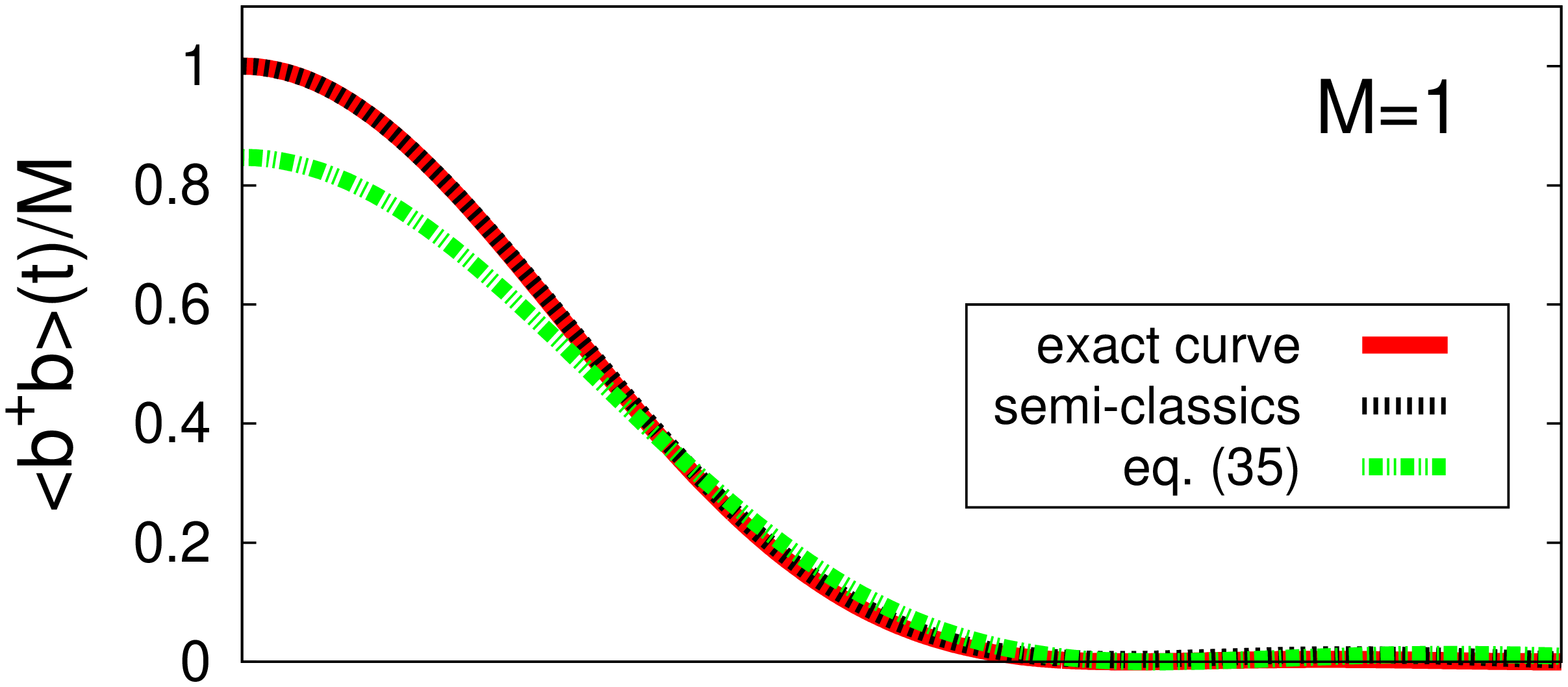}\vspace{-1.4cm}
      \includegraphics[width=0.27\textwidth, angle=270]{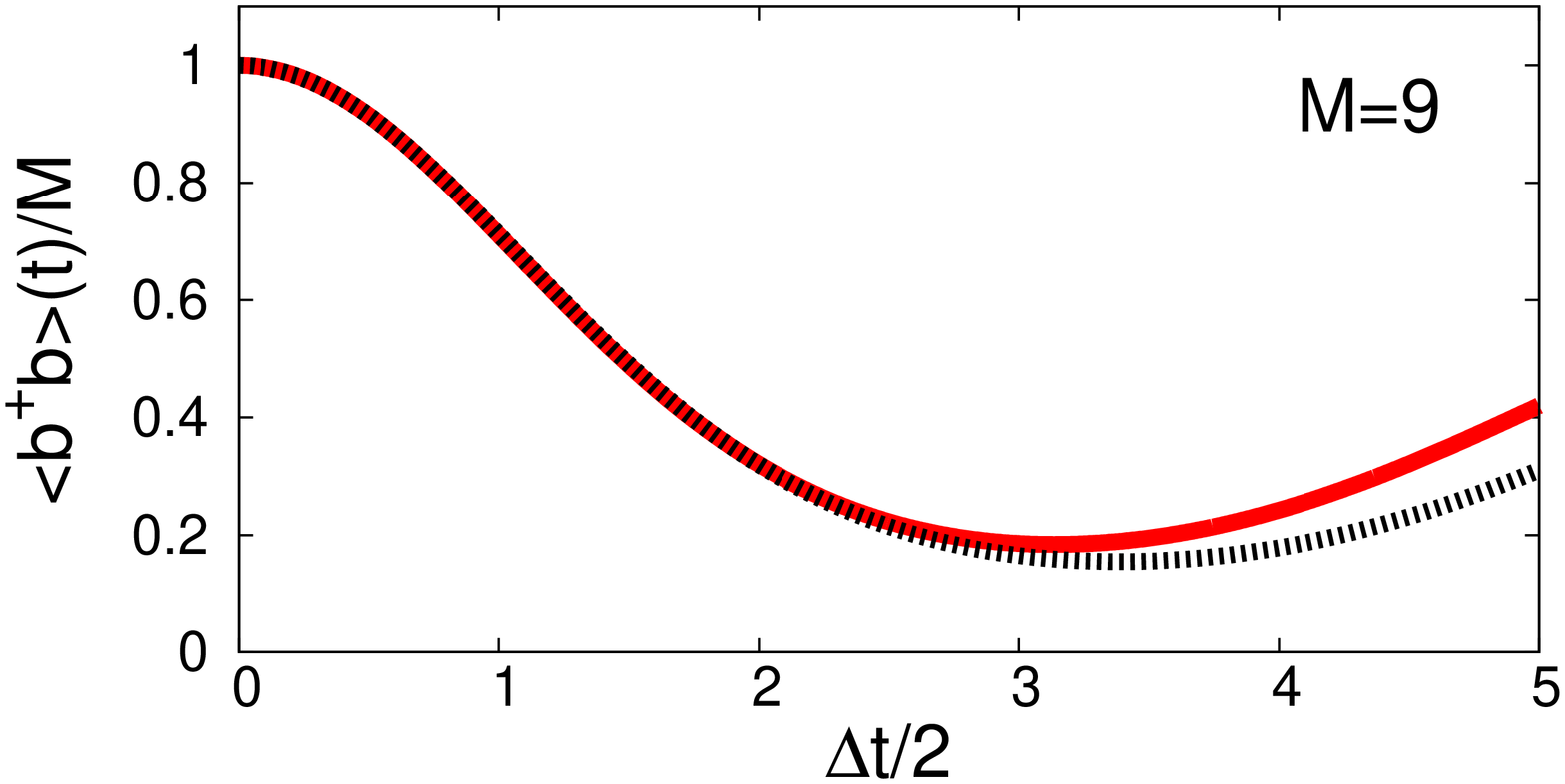}
  \caption{Time evolution of bosonic occupation number, exact (red solid) and meanfield (black dotted) for $N=12$ spins and $M=1,9$ excitations. For $M=1$, the decay part of eq. (\ref{decay}) is shown (green dotted-dashed). Coupling strength is $\Omega/\Delta=0.3$.}
  \label{dyn12}
\end{figure}

Considering that the non-universal behavior of this particular regime is characterized by only a few effective degrees of freedom regardless of the number of excitations, we move away to Rabi frequencies larger than the level spacing but still significantly smaller than the total band width, $\left( \Delta / N \right) < \Omega < \Delta$. Since more and more spins get significantly mixed with the bosonic mode, this ultimately leads to the ``universal weak-field regime'' which exhibits a decay of the bosonic population. However, in this particular regime, any truncation of the Hilbert space leads to an important loss of information and therefore to a large error evidenced by a badly saturated sum rule (\ref{sumrule}). In trying to address the validity of the mean-field approach at such Rabi frequencies we therefore have to limit ourselves to small system sizes. In fact, the number of eigenstates has to be small enough to be able to compute every one of them in a reasonable amount of time therefore accessing exact results. Fig. \ref{dyn12} presents a comparison of the mean-field and the exact quantum dynamics for a small system containing only $N=12$ spin degrees of freedom.

In the presence of a single excitation $M=1$, the bosonic occupation number $\langle b^{\dag}b\rangle(t)$ rapidly decays to almost $0$, which is remarkably well captured by the solution of the semi-classical Eqs. (\ref{eq:QEOM2}). The origin of  this decay lies in the significant overlap of many eigenstates of the system with the initial state, in contrast to the limited number of important eigenstates in the $\Omega \gg \left(\Delta/N \right)$ regime. Decomposing the dynamics into a persistent oscillation [Eq. (\ref{eq:harmonic}] and a decay part,  
\begin{equation}
\sqrt{\langle b^\dagger b\rangle (t) }=\int_{0}^{1}\frac{\left(4\Omega^{2}/\Delta^{2}\right)\cos\left(y\Delta t/2\right) dy}{\left(y-\frac{2\Omega^{2}}{\Delta^{2}}\ln\left(\frac{1+y}{1-y}\right)\right)^{2}+\left(\frac{2\pi\Omega^{2}}{\Delta^{2}}\right)^{2}},
\label{decay}
\end{equation}
as in Ref. \onlinecite{oleks_dynamics}, we indeed find that  the the continuum part of the energy spectrum is dominant.
 Moving away from $M\ll N$ we look at a strongly excited initial state containing $M=9$ bosons. Once again the initial decay is perfectly reproduced by the mean-field treatment in spite of the relatively large number of excitations. At later times the exact quantum treatment deviates from the mean-field approach due to small size of the system.

\begin{figure}
      \includegraphics[width=0.275\textwidth, angle=270]{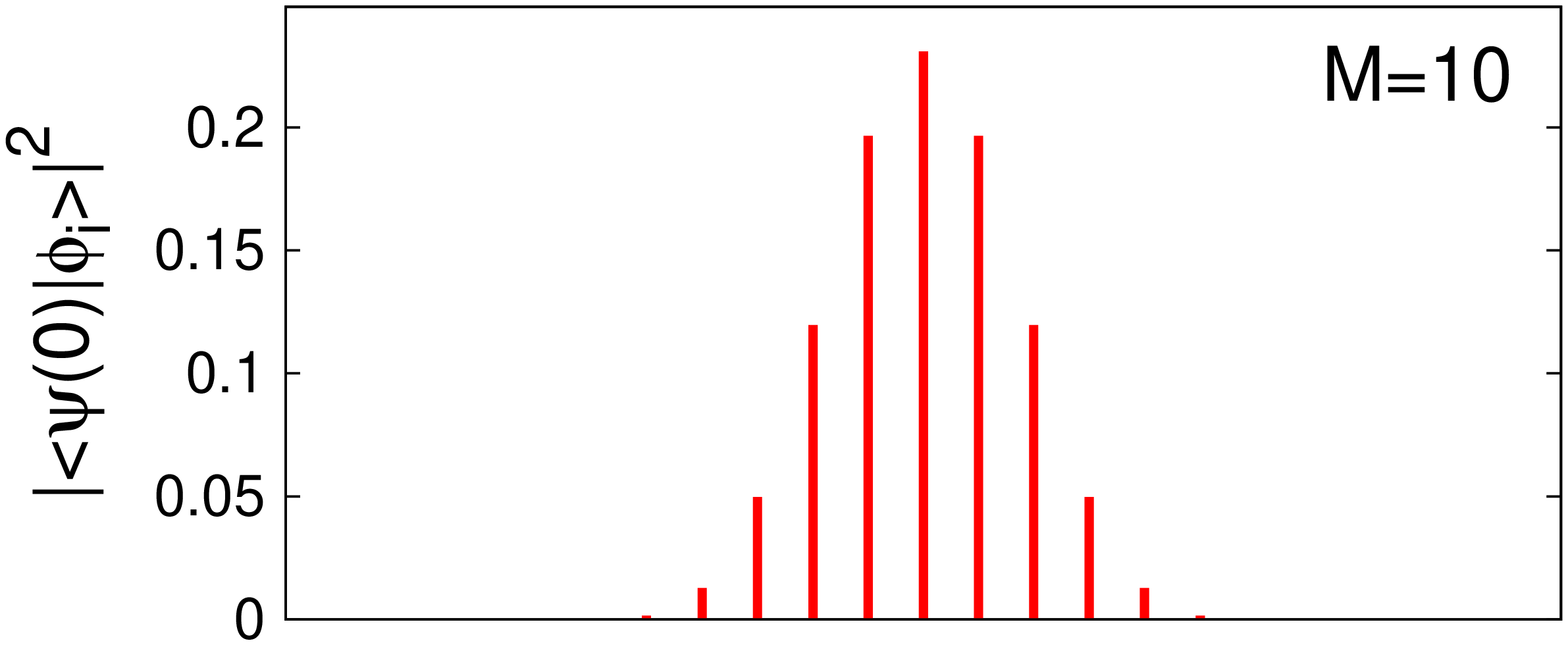}\vspace{-1.67cm}
      \includegraphics[width=0.275\textwidth, angle=270]{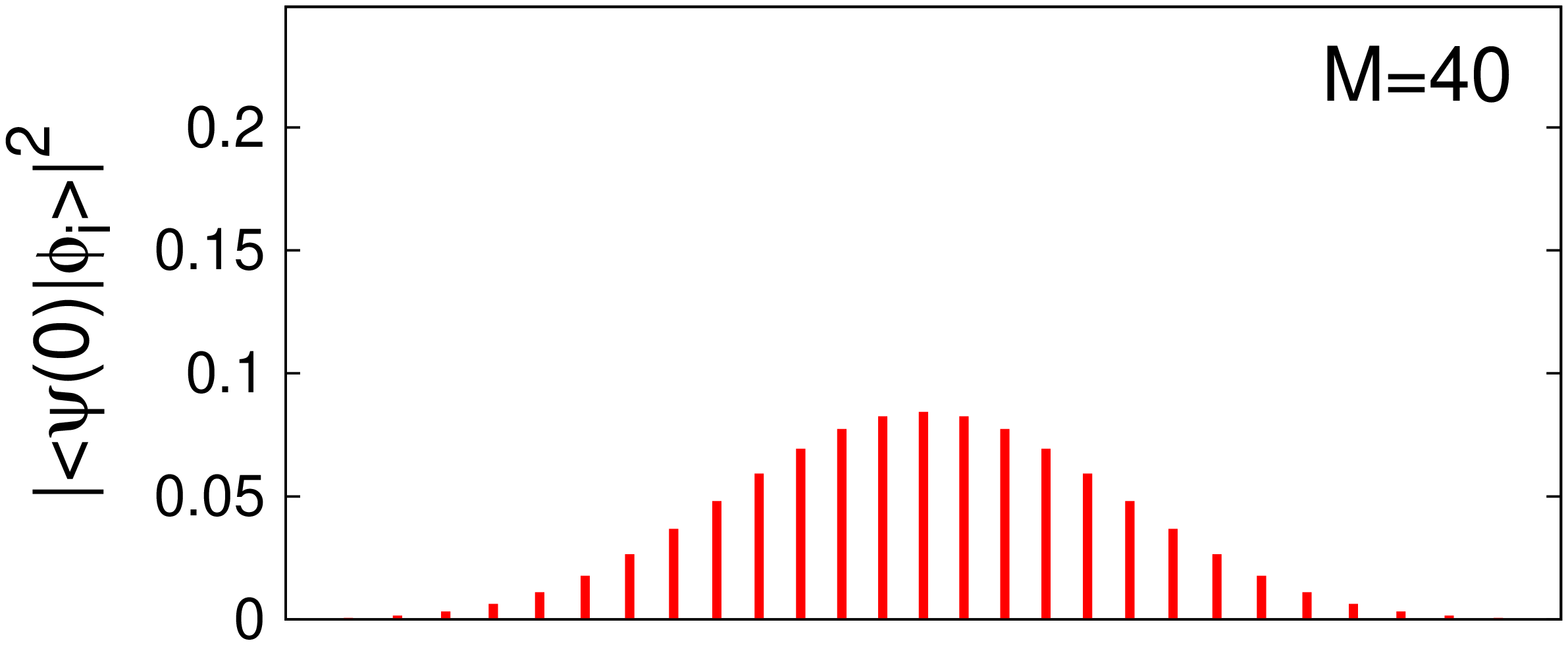}\vspace{-1.67cm}
      \includegraphics[width=0.275\textwidth, angle=270]{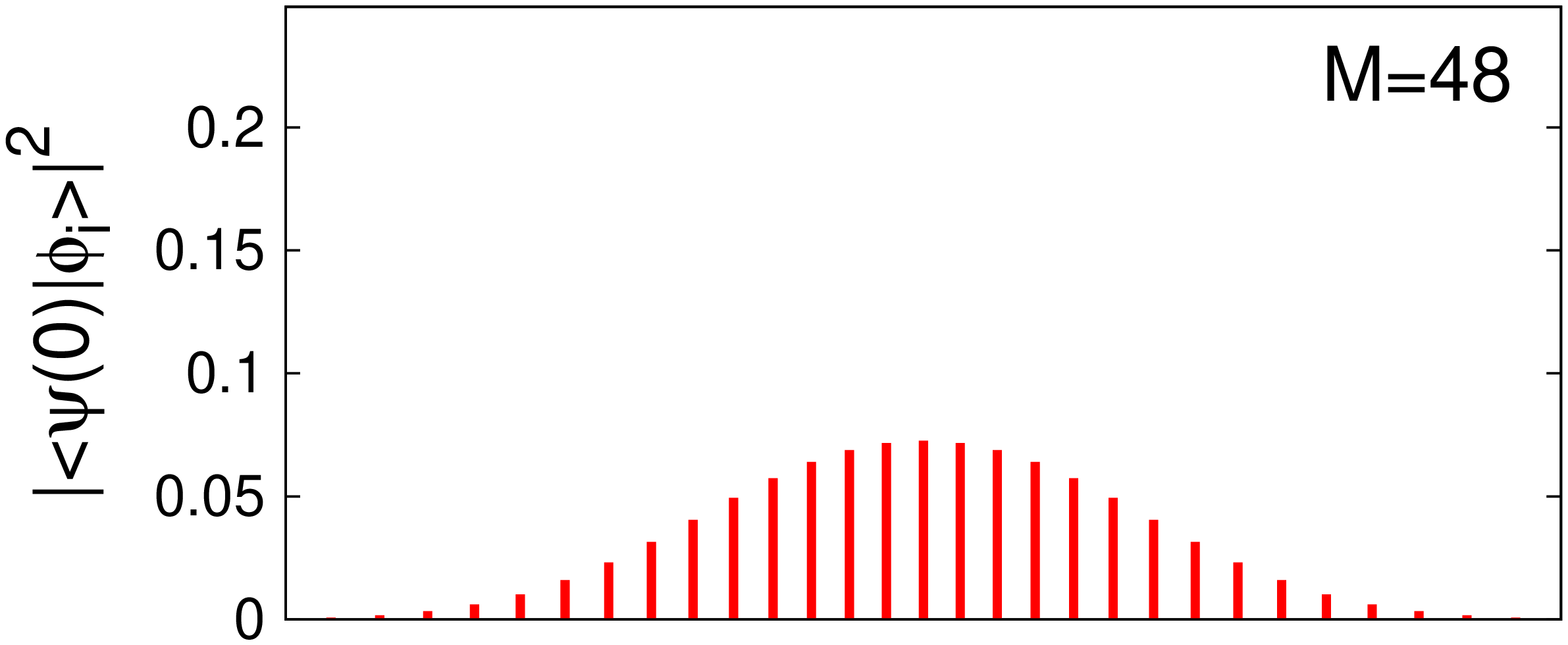}\vspace{-1.68cm}
      \includegraphics[width=0.275\textwidth, angle=270]{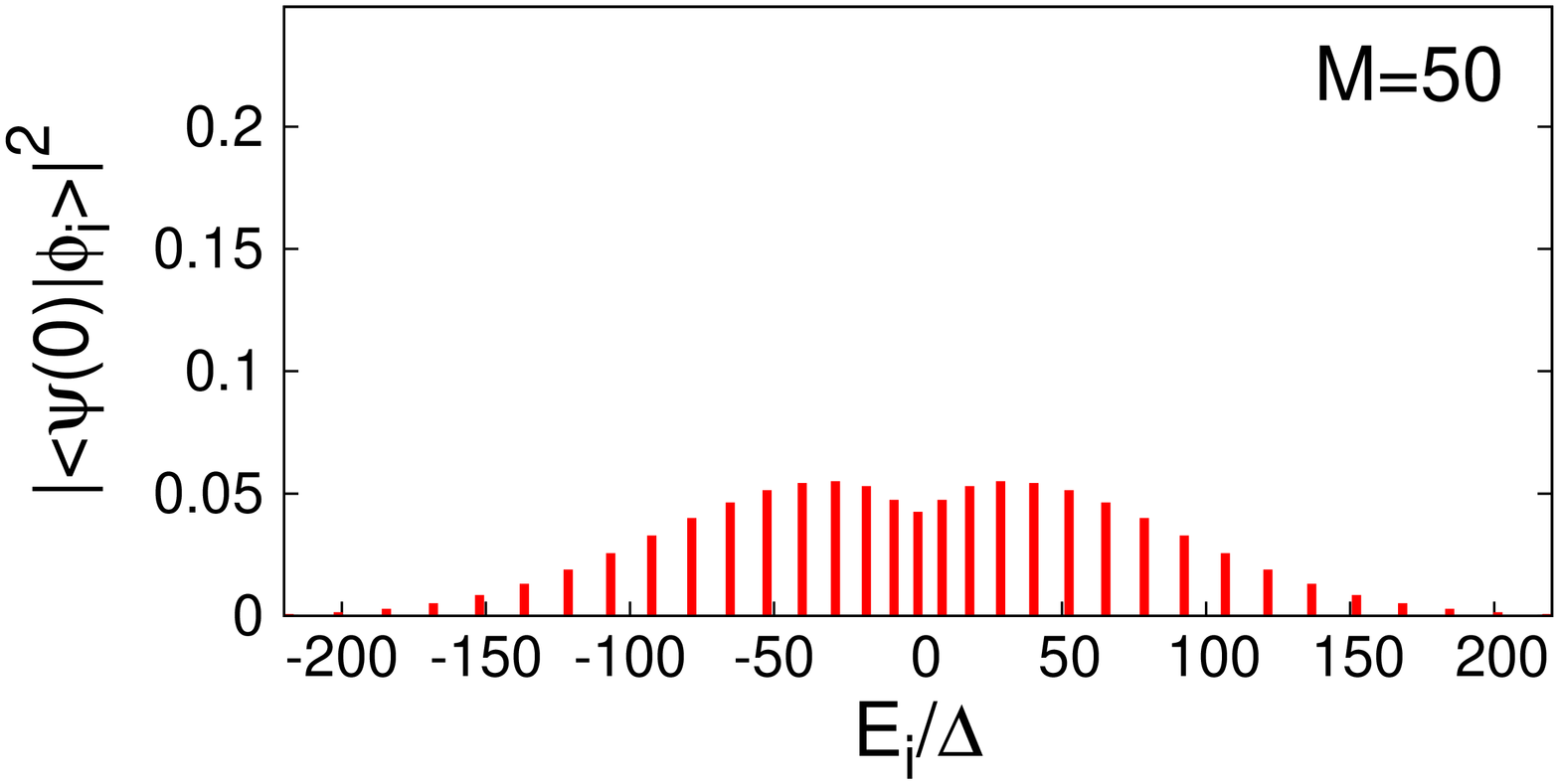}
  \caption{Work distribution function for $N=50$ spins and $M=10,40,48,50$ exctitations. Coupling strength is $\Omega/\Delta=10$.}
  \label{fig:freq_spectrum}
\end{figure}

Within the restriction to short times imposed by finite size effects,  the mean-field approach remains valid even at  large excitation numbers in this regime.  However, the severe limitation on the system size makes it difficult to reach a precise conclusion for any larger systems. We therefore turn our focus to the strong coupling regime, $\Omega >\Delta$. In this limit, a drastic truncation becomes possible while maintaining a satisfying saturation of the sum rule (\ref{sumrule}). We first present in Fig. \ref{fig:freq_spectrum}, the energy spectrum characterizing the time evolution of the bosonic occupation numbers for a system of $N=50$ spins. Specifically, we plot the work distribution $P(E)=\sum_{i=1}^d \left|\braket{M;\down\ldots\down}{\phi_i}\right|^2 \delta(E-E_i)$ which, according to eq. (\ref{bosonocc}), describes the frequency content of the initial condition. For these calculations the dimension of the Hilbert space is reduced from $d= O(10^{15})$ (for $M=N$) to $\tilde{d}= N+1$ by keeping only the states with $M$ divergent rapidities. Doing so maintains the truncation error in $\langle b^{\dag}b\rangle(t)/M$ below a maximum of $\delta=5\%$.   

For low excitation numbers, the spectrum presents itself as a series of nearly equally spaced peaks. According to Eq. (\ref{bosonocc}), this constant energy difference between the contributing eigenstates indicates a periodic oscillation in the time evolution.
However, as the number of excitations is increased, deviations from the harmonic progression become more and more important and is particularly evident in the $M=N$ results where the low energy contributions are much closer than the high energy ones. Additionally, the shape of the distribution is severely altered. 
Therefore, while one can expect the  periodicity of the semi-classical results to be adequately reproduced for small $M\ll N$, the opposite regime will be characterized by a set of incommensurate frequencies ultimately leading to some decay. 

\begin{figure}[h]
      \includegraphics[width=0.275\textwidth, angle=270]{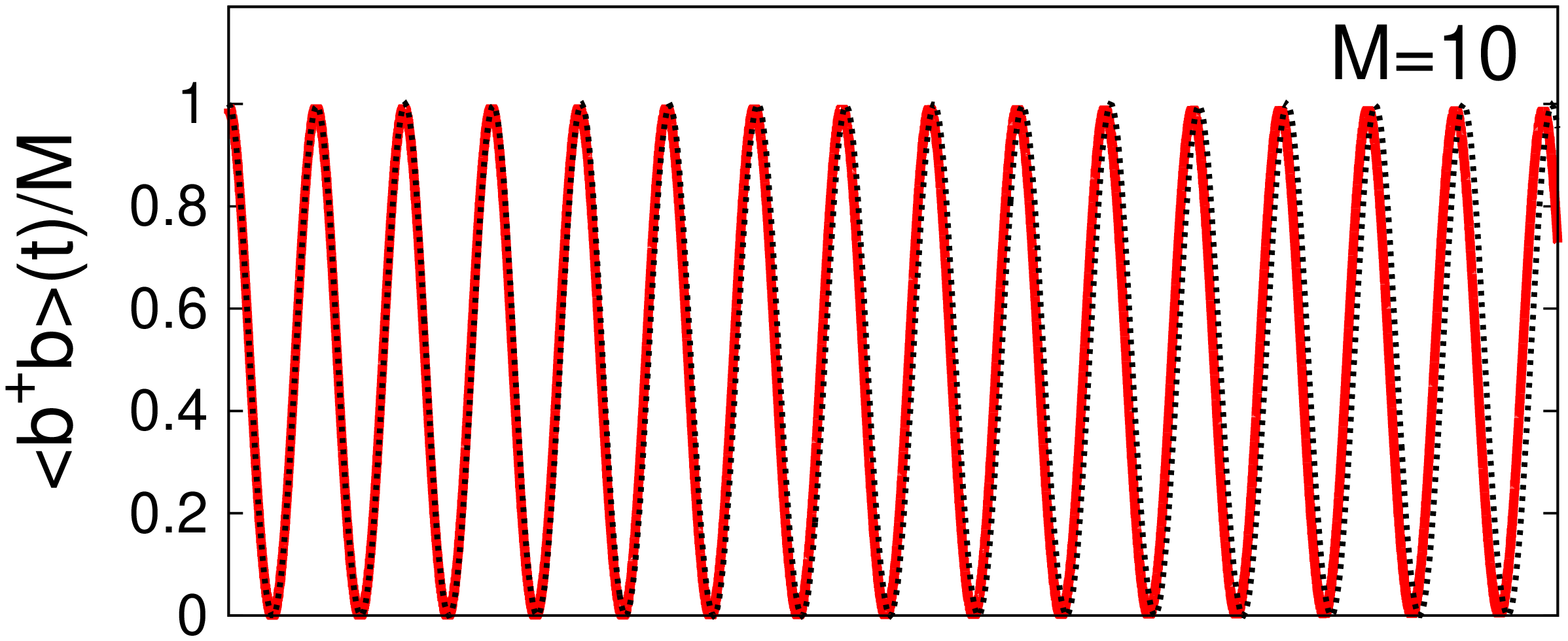}\vspace{-1.67cm}
      \includegraphics[width=0.275\textwidth, angle=270]{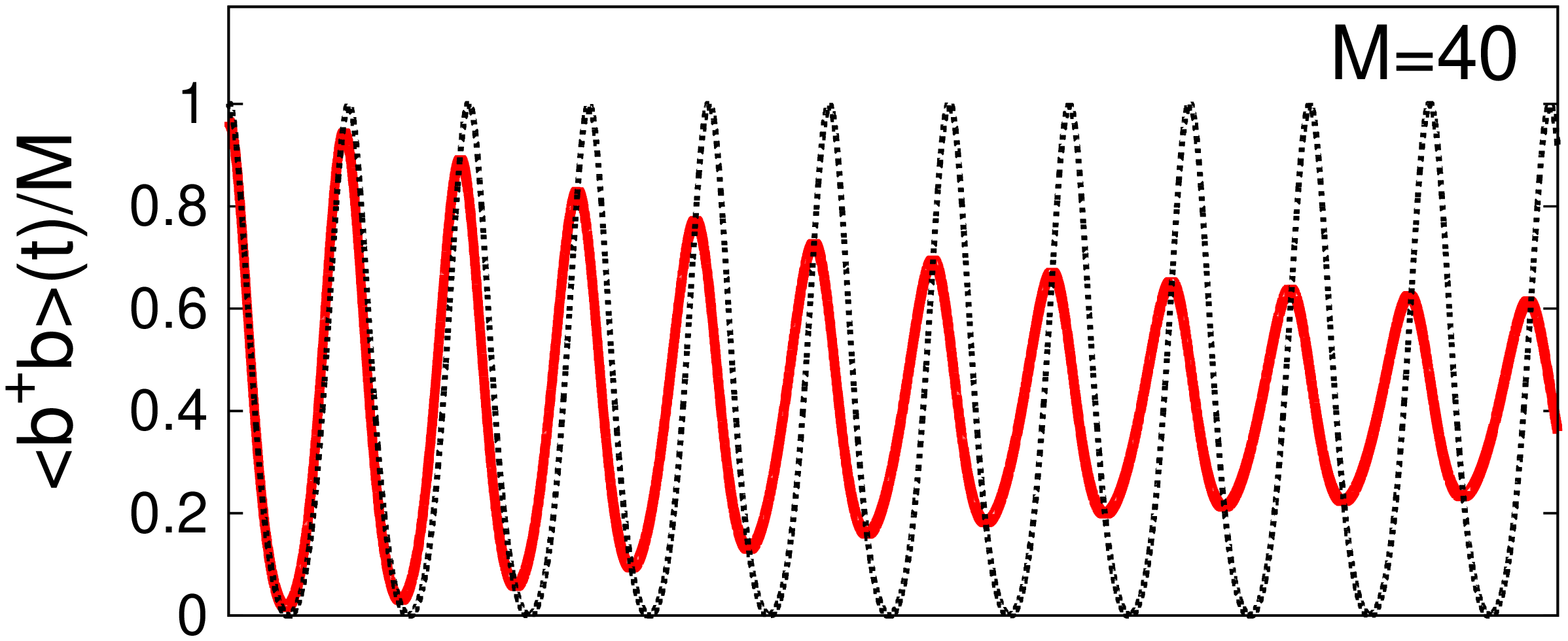}\vspace{-1.67cm}
      \includegraphics[width=0.275\textwidth, angle=270]{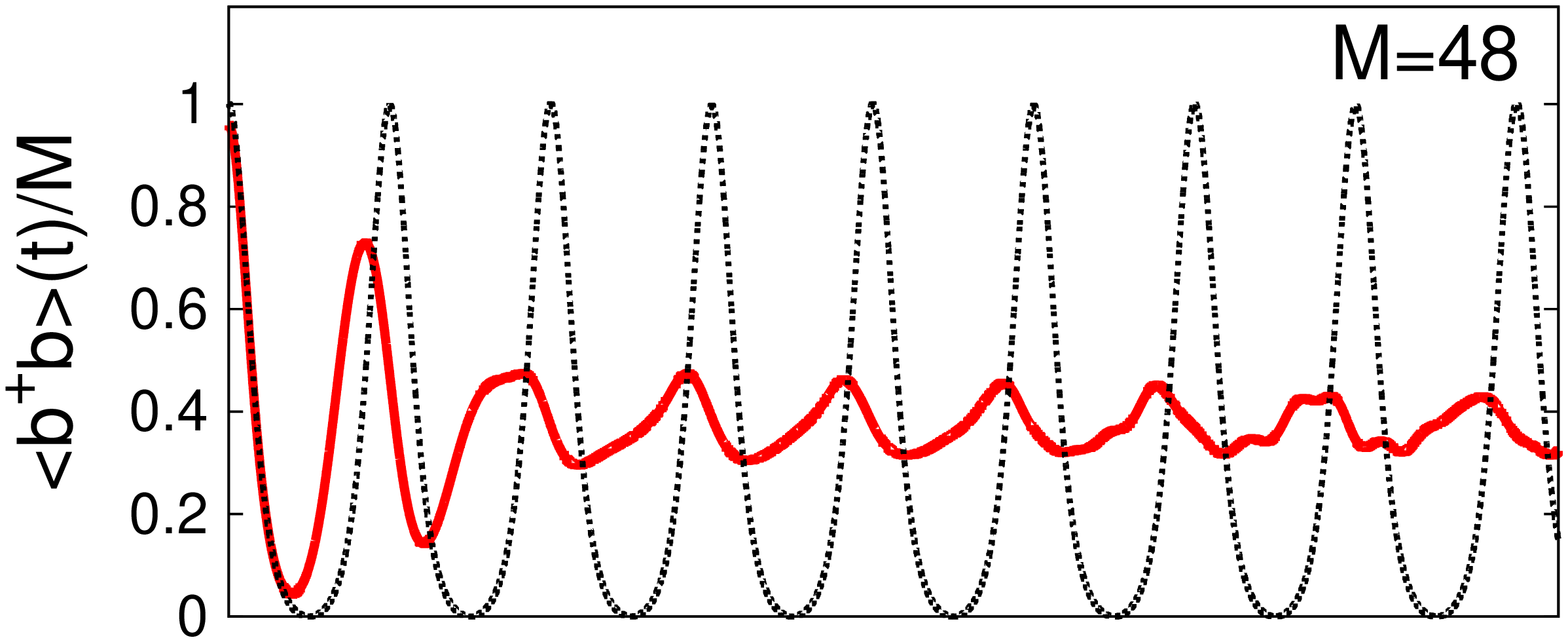}\vspace{-1.67cm}
      \includegraphics[width=0.275\textwidth, angle=270]{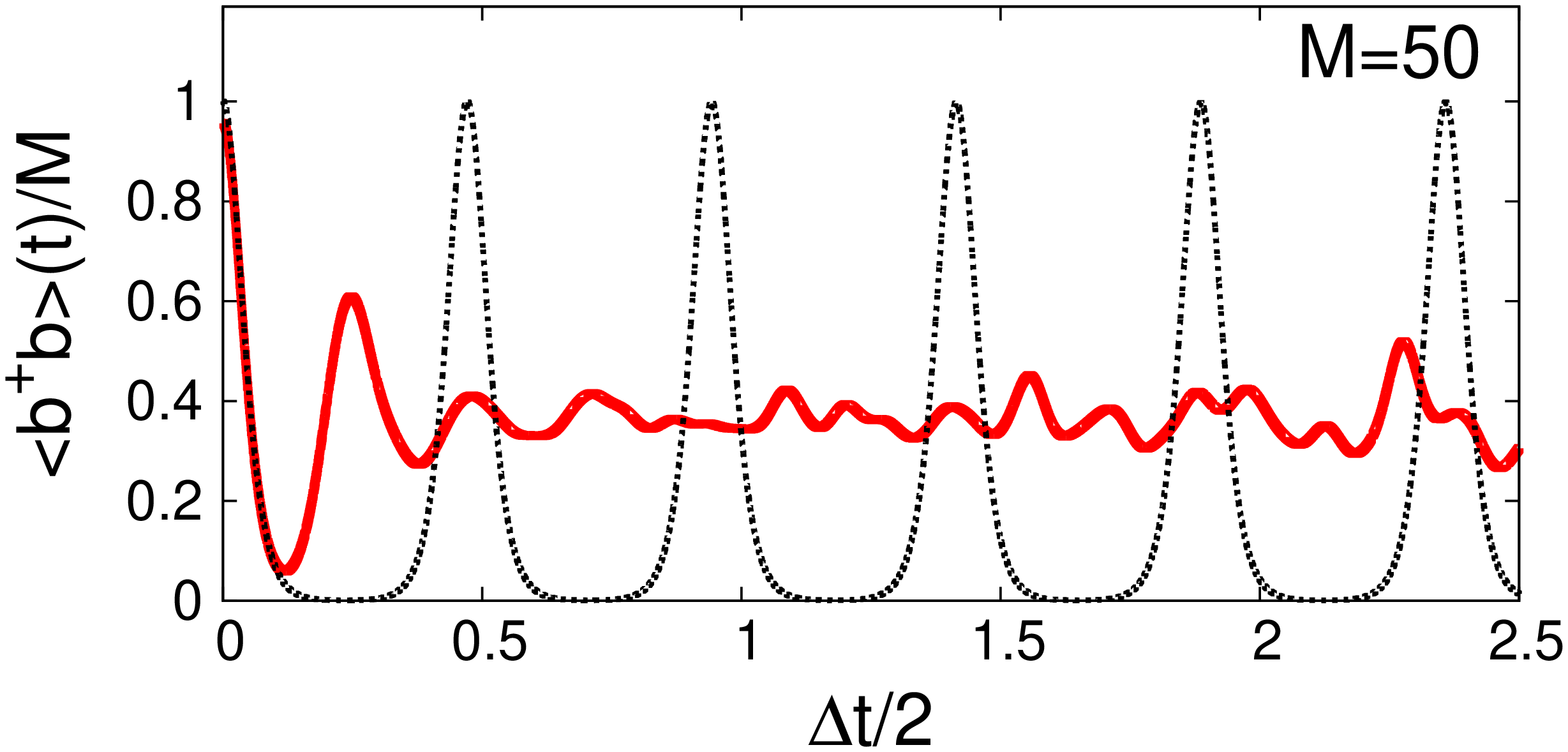}
  \caption{Time evolution of bosonic occupation number, exact (red solid) and mean-field (black dotted) for $N=50$ spins $M=10,40,48,50$ and $\Omega/\Delta=10$.}
  \label{fig:time_evolution}
\end{figure}

This is evidenced by looking at the explicit time evolution of the average bosonic occupation presented in Fig. \ref{fig:time_evolution} for a variety of initial number of excitations. Both the mean-field behavior (black) and the quantum evolution (red) are plotted for $N=50$ spins.

Two main differences between the quantum and mean-field dynamics are seen. First, the quantum oscillation frequency is systematically shifted to higher frequencies and, at the same time, the amplitude is shown to decay. When the number of excitations is small enough the frequency shift is small and the decay is slow compared to the time scale set by the oscillation frequency. However, when $M$ and $N$ become comparable, the quantum dynamics exhibits a rapid amplitude decay which, not being captured by the mean-field analysis, leads to a drastic difference between both descriptions of the bosonic occupation. As evidenced by Fig. \ref{fig:time_evolution}, a larger difference in the oscillation frequency also occurs, making the distinction between both approaches even more important. Nonetheless, even for $M=N$ the mean-field approach is shown to capture perfectly the initial instability and provides an accurate description up to some finite time.

\begin{figure}[h]
  \centering
     \includegraphics[width=0.33\textwidth, angle=270]{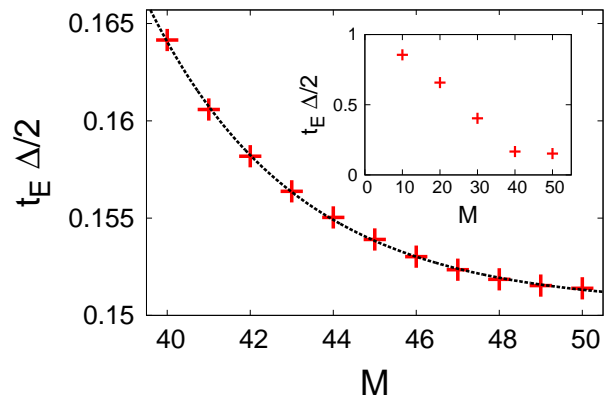}
  \caption{Outer plot: Ehrenfest time for N=50 as a function of excitation number $M$ for high filling factors. Dashed line is a $t_e(M) = (t_\infty+b e^{-c M})$ fit. Inset: The same plot for the whole range of $M$.}
  \label{fig:ehrenfest_time}
\end{figure}

In order to characterize the regime of validity of the mean-field approximation, we extract an Ehrenfest time by looking for the earliest time where the mean-field and quantum bosonic occupations differ by $10\%$ of the initial population. The resulting times are plotted as a function of the excitation number in Fig. \ref{fig:ehrenfest_time}. 

For few excitations $M\ll N$ both the mean-field and quantum numerical calculations were shown to coincide up to $1/N$ corrections \cite{oleks_dynamics}. Here we see that the Ehrenfest time initially undergoes a rapid, seemingly linear, decrease as $M$ increases. When the strongly excited regime is reached, at $M\approx 0.8 N$, this behavior is drastically modified. It is then well described by an exponential fit $t_e(M) = (t_\infty+b e^{-c M})$ with parameters $t_\infty=0.15,b=1003.65,c=0.28$. The saturating decrease of the Ehrenfest time indicates that the mean-field description retains its validity in the description of the initial stages even as $M$ reaches $N$.

\begin{figure}[h]
  \centering
     \includegraphics[width=0.32\textwidth, angle=270]{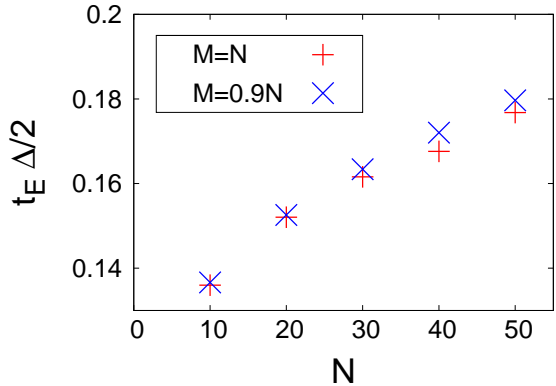}
  \caption{Ehrenfest time for $M=0.9N$ and $M=N$ as a function of the system size $N$ for $\Omega/\Delta=10$.}
  \label{fig:ehrenfest_time_size}
\end{figure}

For a low number of initial excitations \cite{oleks_dynamics}, the mean-field approximation is know to be exact in the limit $N\to \infty$. Moreover, as evidenced in Ref. [\onlinecite{oleks_dynamics}] and in this work, a modest mesoscopic number of spins $N\approx\mathcal{O}(10^2)$ is sufficient for the mean-field treatment to adequately describe the dynamics over many oscillation periods. For a strongly excited system, we plot in Fig. \ref{fig:ehrenfest_time_size} the system size dependence of the Ehrenfest time obtained at large fillings $(M=N,  M = 0.9N)$. Even for a large number of excitations, the growth of $t_E$ with increasing system size indicates that the mean-field approach could still provide an adequate description, even at long times, for thermodynamically large $N\to \infty$ systems. However, since  this growth is extremely slow and therefore in stark contrast to the $M \ll N$ case, mesoscopic systems remain too small for the classical mean-field treatment to describe correctly the behavior past the initial stage of decay.

\section{Conclusions}

By exploiting the quantum integrability of the Dicke model, we were able to calculate the quantum dynamics of an initially populated single bosonic mode interacting with an ensemble of inhomogeneous ensemble of two-level systems. For strong enough couplings, this method based on the algebraic Bethe ansatz provides a simple truncation scheme, which allowed us to treat relatively large systems even for a strongly excited initial state.

We compared the numerical solutions of its non-equilibrium dynamics with its mean-field description. Focusing on the strong coupling regime, where mean-field theory predicts oscillating periodic solutions, we confirm that at low excitation numbers both solutions agree up to a relatively long finite Ehrenfest time. However, going to more strongly excited systems leads to a rapid shortening of the mean-field description's period of validity due to a shift in the oscillation frequency combined with a decay of the oscillation's amplitude which are exclusively captured by a full quantum treatment . For relatively large mesoscopic systems, we demonstrate that, although it cannot capture the long time dynamics, the initial decay of the bosonic excitations is still adequately described by the classical mean-field theory. 

\section*{Acknowledgments}

A.F.'s and C.S.'s work was supported by the DFG through SFB631, SFB-TR12 and the Excellence Cluster Nanosystems Initiative Munich (NIM).  O.T.'s work was supported by the EPSRC (UK) EP/G001642.



\begin{thebibliography}{999}

\bibitem{qucom1}

 J. I. Cirac and P. Zoller, Phys. Rev. Lett. {\bf74}, 4091 (1995).

\bibitem{qucom2}
A. Imamo\u{g}lu, D. D. Awschalom, G. Burkard, D. P. Di-Vincenzo, D. Loss, M. Sherwin, and A. Small, Phys. Rev. Lett. {\bf83}, 4204 (1999).

\bibitem{qucom3}
  L. Childress, A. S. Srensen, and M. D. Lukin, Phys. Rev. A {\bf69}, 042302 (2004).

\bibitem{qucom4}

 A. Wallraff, D. I. Schuster, A. Blais, L. Frunzio, R.- S. Huang, J. Majer, S. Kumar, S. M. Girvin, and  R. J. Schoelkopf, Nature {\bf 431}, 162 (2004).

\bibitem{excitons1}

J. Kasprzak, M. Richard, S. Kundermann, A. Baas, P. Jeambrun, J. M. J. Keeling, F. M. Marchetti, M. H. Szymanska, R. Andr\'{e}, J. L. Staehli, V. Savona, P. B. Littlewood, B. Deveaud and Le Si Dang, Nature {\bf 443}, 409 (2006).

\bibitem{excitons2}
R. Balili, V. Hartwell, D. Snoke, L. Pfeiffer, K. West, Science {\bf  316}, 1007 (2007).

\bibitem{eastham}
P. R. Eastham and P. B. Littlewood, Phys. Rev. B 64, 235101 (2001).

\bibitem{dots1}

J. Berezovsky, M. H. Mikkelsen, N. G. Stoltz, L. A. Coldren, and D. D. Awschalom, Science {\bf 320}, 349 (2008)

\bibitem{dots2} K. Hennessy, A. Badolato, M. Winger, D. Gerace, M. Atat\"{u}re, S. Gulde, S. F\"{a}lt, E. L. Hu and A. Imamo\u{g}lu, Nature {\bf 445}, 896 (2007).

\bibitem{phillips} Yanwen Wu, I. M. Piper, M. Ediger, P. Brereton, E. R. Schmidgall, P. R. Eastham, M. Hugues, M. Hopkinson, and R. T. Phillips,  Phys. Rev. Lett. {\bf 106}, 067401 (2011).

\bibitem{QD_lasers1}
L. Harris, A. D. Ashmore, D. J. Mowbray, M. S. Skolnick, M. Hopkinson, G. Hill, and J. Clark, Appl. Phys. Lett. {\bf 75}, 3512 (1999).

\bibitem{QD_lasers2}

H. C. Schneider, W. W. Chow, and S. W. Koch, Phys. Rev. B {\bf 66}, 041310(R) (2002).
\bibitem{QD_lasers3}

 D. Goulding, S. P. Hegarty, O. Rasskazov, S. Melnik, M. Hartnett G. Greene, J. G. McInerney, D. Rachinskii, and G. Huyet, Phys. Rev. Lett. {\bf 98}, 153903 (2007).

\bibitem{QD_lasers4}
T. Erneux, E. A. Viktorov, B. Kelleher, D. Goulding, S. P. Hegarty, and G. Huyet, Optics Letters {\bf 35}, 937 (2010).

\bibitem{arakawa}
Y. Arakawa and H. Sakaki, Appl. Phys. Lett. {\bf 40}, 939 (1982).

\bibitem{dicke}
 R. H. Dicke, Phys. Rev. {\bf 93}, 99 (1954).

\bibitem{nonRWA} Here we use the rotating wave approximation assuming that $V_j\ll\epsilon_j$. A study of the light-matter coupling without this approximation, e.g. effects of quantum chaos in level-statistics,\cite{EmeryBrandes} is outside the scope of our work.

\bibitem{EmeryBrandes}
C. Emary and T. Brandes, Phys. Rev. Lett. {\bf 90}, 044101 (2003).

\bibitem{oleks_dynamics}
 O. Tsyplyatyev and D. Loss, Phys. Rev. A {\bf 80}, 023803 (2009).

\bibitem{oleks_ehrenfest}
 O.Tsyplyatyev, D. Loss, Phys. Rev. B {\bf 82}, 024305 (2010).
 
\bibitem{Vj_many}  When M is comparable with N, a semiclassical analysis of Eq. (1) in Ref. \onlinecite{Eastham_semiclassics} suggests that $V_j\neq V$ may introduce some additional effects.

\bibitem{Eastham_semiclassics}P. R. Eastham, J. Phys.: Condens. Matter {\bf 19}, 295210 (2007).

\bibitem{HeppLieb} K. Hepp and E. H. Lieb, Ann. Phys. (N.Y.) {\bf 76}, 360 (1973).

\bibitem{InGaAsdots} K. L. Silverman, R. P. Mirin, S. T. Cundiff, and A. G. Norman, Appl. Phys. Lett. {\bf 82}, 4552 (2003).

\bibitem{Lemaitre} V. Loo, L. Lanco, A. Lemaitre, I. Sagnes, O. Krebs, P. Voisin, and P. Senellart, Appl. Phys. Lett. {\bf 97}, 241110 (2010).

\bibitem{Warburton} R. J. Barbour, P. A. Dalgarno, A. Curran, K. M. Nowak, H. J. Baker, D. R. Hall, N. G. Stoltz, P. M. Petroff, and R. J. Warburton, J. Appl. Phys. {\bf 110}, 053107 (2011).

\bibitem{Nestimate} We assume that a single layer of quantum dots is put inside a cavity whose linear dimension is $2.8$ $\mu$m as in Ref. \onlinecite{Warburton}.

\bibitem{bandwidth} N. Perret, D. Morris, L. Franchomme-Foss\'{e}, R. C\^{o}t\'{e}, S. Fafard, V. Aimez, and J. Beauvais, Phys. Rev. B {\bf 62}, 5092 (2000). 

\bibitem{faribaultjmp}
A. Faribault, P. Calabrese and J.-S. Caux, J. Math. Phys. {\bf 50}, 095212 (2009).

\bibitem{methods}
A. Faribault, O. El Araby, C. Str\"{a}ter and V. Gritsev, Phys. Rev. B {\bf 83}, 235124 (2011).

\bibitem{methods2}
 O. El Araby, V. Gritsev and A. Faribault  Phys. Rev. B {\bf 85}, 115130 (2012).


\bibitem{poorman}
O. Tsyplyatyev, J. von Delft and D. Loss, Phys. Rev. B {\bf 82}, 092203 (2010).

\bibitem{slavnov}
N. A. Slavnov, Theoretical and Mathematical Physics {\bf 79} 502-508 (1989).

\bibitem{links}
J. Links, H. Zhou, R. H. McKenzie and M. D. Gold, J. Phys. A: Math. {\bf 36} R63 (2003).




\bibitem{fermi_bose}
E. A. Yuzbashyan, V. B. Kuznetsov and B. L. Altshuler , Phys. Rev B {\bf 72}, 144524 (2005).

\bibitem{relaxation}
E. A. Yuzbashyan, O. Tsyplyatyev, and B. L. Altshuler, Phys. Rev. Let. {\bf 96}, 097005 (2006).

\bibitem{dicke_proof}
J. Dukelsky, G. G. Dussel, C. Esebbag, and S. Pittel, Phys. Rev. Let., {\bf 93}, 050403 (2004).

\end{thebibliography}
\end{document}